\documentclass[10pt,twocolumn]{article}
\usepackage{times}
\usepackage{graphicx}
\usepackage{hyperref}
\usepackage{amsmath}
\usepackage{amssymb}
\usepackage{booktabs}
\usepackage{enumitem}
\usepackage[numbers,sort&compress]{natbib}
\usepackage[margin=1in]{geometry}
\usepackage{abstract}
\usepackage{tikz}
\usetikzlibrary{
    positioning,
    shapes.geometric,
    arrows.meta,
    fit,
    backgrounds,
    calc
}
\usepackage{xcolor}        
\usepackage{orcidlink}     

\tikzset{
    block/.style={rectangle, rounded corners, draw=blue!70!black, fill=blue!10, thick, minimum width=3.2cm, minimum height=1.2cm, align=center, font=\sffamily\bfseries\small},
    cloud_service/.style={rectangle, rounded corners, draw=orange!90!black, fill=orange!12, thick, minimum width=3.1cm, minimum height=1.1cm, align=center, font=\sffamily\small},
    data_store/.style={cylinder, shape border rotate=90, draw=purple!75!black, fill=purple!10, thick, minimum height=1.4cm, minimum width=2.8cm, align=center, font=\sffamily\small},
    user_node/.style={ellipse, draw=green!60!black, fill=green!10, thick, minimum width=2.9cm, minimum height=1.2cm, align=center, font=\sffamily\small},
    table/.style={rectangle, draw=blue!50!black, fill=blue!10, thick, minimum width=3.2cm, minimum height=1cm, align=center, font=\sffamily\footnotesize},
    group_box/.style={rectangle, rounded corners, draw=gray!50, fill=gray!5, very thick, inner ysep=12pt, inner xsep=10pt},
    arrow/.style={-{Latex[length=2.8mm]}, thick, rounded corners=3pt, color=black!80},
    dashed_arrow/.style={-{Latex[length=2.8mm]}, thick, dashed, rounded corners=3pt, color=black!50},
    rel/.style={-{Latex[length=2.3mm]}, thick, blue!60}
}

\title{\bf Real-Time AI-Driven Pipeline for Automated Medical Study Content Generation in Low-Resource Settings: A Kenyan Case Study}
\author{
  Dr. Emmanuel Korir\,\orcidlink{0009-0006-2739-9535}\\
  \small University of Nairobi, Kenya\\
  \small \texttt{emmanuelqkorir@uonbi.ac.ke}
  \and
    Dr. Eugene Wechuli\\
  \small University of Nairobi, Kenya\\
  \small \texttt{wechuli@juvenotes.com}
}

\date{}

\begin{document}
\twocolumn[
\maketitle
\vspace{-2em}

\begin{onecolabstract}
\noindent
\textbf{Background:} Medical education in low- and middle-income countries faces critical bottlenecks in content creation, infrastructure, and connectivity, contributing to projected shortages of 6.1\,million health workers in Africa by 2030. Leveraging advances in AI, particularly large language models (LLMs) and document-analysis services, offers a pathway to automate and scale study material generation.

\textbf{Objective:} This study evaluates \emph{Juvenotes}, a real-time AI-driven pipeline that transforms academic documents into structured exam-style question banks, with UX optimizations for low-resource settings in Kenya.

\textbf{Methods:} Juvenotes integrates Azure Document Intelligence for OCR and Azure AI Foundry (OpenAI o3-mini) for question/answer generation within a microservices architecture deployed on Azure VMs. A Vue/TypeScript frontend and AdonisJS backend handle user interactions, while the Python-based AI engine orchestrates document processing, LLM prompting, and database insertion. The system includes mobile-first design, bandwidth-sensitive interfaces, institutional tagging, and offline-capable features. Over seven months, the platform was piloted at five Kenyan institutions, and evaluation metrics included daily active users, content processing time, and user satisfaction surveys.

\textbf{Results:} Deployment of Juvenotes reduced content curation time from days to minutes and generated high-quality question banks in real time. Daily active user counts rose by 40\%, and 90\% of surveyed students reported improved study experiences compared to traditional methods. Key challenges included intermittent connectivity, power disruptions, and occasional AI-generated errors, which highlighted the necessity for offline synchronization, human-in-the-loop validation, and model fine-tuning with local curricula.

\textbf{Conclusions:} Juvenotes demonstrates that AI-powered automation, when combined with context-aware UX adaptations, can significantly enhance access to quality study materials in low-resource medical education settings. Future work should focus on expanding offline capabilities, integrating rigorous evaluation frameworks, and securing long-term institutional support to sustain impact and help address Africa’s health-worker shortfall.
\end{onecolabstract}
\vspace{1em}
]

\section{Introduction}
Medical education in low- and middle-income countries, particularly in Africa, is faced with a severe shortage of healthcare workers~\cite{who2025staffshortfalls,zyl2021unravelling, barteit2020evaluation}. Sustainable Development Goals (SDGs) related to health and quality education have also not been attained in most African countries~\cite{barteit2020evaluation}. Currently, the medical educational infrastructure is inadequate and faces a shortage of medical educators and learning facilities~\cite{zyl2021unravelling}. The World Health Organization (WHO) projects a regional shortage of approximately 6.1 million health workers in Africa by 2030~\cite{who2025staffshortfalls}. E-learning has been proclaimed as a revolutionary force in enhancing medical education across Africa. It has been proposed to eliminate medical education barriers without the need for extensive infrastructure investment.

Despite this promised potential, a major hurdle in LMIC has been the manual creation of high-quality study content, infrastructural limitations, and connectivity issues~\cite{barteit2020evaluation}. Large parts of Africa still struggle with poor internet connectivity, high costs of mobile data, and lack of electricity. According to the World Bank Group~\cite{barteit2019elearning,worldbank2023digital}, only 36\% of Africans have access to broadband internet. This often leads to small-scale e-learning pilot projects that fail to scale beyond the pilot phase.

Recent advancements in Artificial Intelligence (AI), specifically Large Language Models (LLMs), provide an opportunity to automate and scale educational content generation. LLMs can generate educational content and personalized feedback, which can enhance medical education and clinical skills training. Based on these advancements, this paper presents Juvenotes, an AI-powered educational platform specifically developed to automate the process of extracting and generating exam-style questions and answers from academic documents. This is particularly important for use in low-resource medical training environments.

Piloted in Kenyan medical schools, Juvenotes leverages Azure cloud AI services, including Azure Document Intelligence for Optical Character Recognition (OCR) and Azure AI Foundry for generating structured exam-style questions and their answers. These services are orchestrated via a containerized microservices architecture, enhancing portability and resilience. The platform’s development has heavily prioritized user experience (UX) adaptations tailored specifically for low-bandwidth environments and the African context. Early pilot data from Kenyan institutions, particularly the University of Nairobi, have demonstrated a notable 40\% rise in student engagement and a drastic reduction in content incorporation time from days to minutes. This affirms the significant potential of AI in bridging critical educational gaps in LMIC.

\section{Related Work}
Automatic question generation (AQG) using LLMs is extensively gaining traction. Zhu et al.~\cite{zhu2024potential} evaluated eight LLMs in generating exam-style items based on medical records. They reported strengths in question formation and weaknesses in answer accuracy and domain specificity. Yao et al.~\cite{yao2024mcqg} introduced a framework (MCQG-SRefine) that combined self-critique and correction loops to enhance multiple choice questions (MCQ) quality, specifically for United States Medical Licensing Examination (USMLE) questions. Previous work, such as Med-PaLM 2, attained physician-level performance in the medical questions datasets~\cite{singhal2023towards}.

E-learning in African medical education often occurs in isolated pilot programs. Barteit et al.~\cite{barteit2019elearning} noted that such interventions tend to underperform due to insufficient technical, institutional, and user support. Makerere University's research during COVID-19 identified poor internet connectivity and high internet costs as the barriers to e-learning~\cite{olum2020medical}. The Medical Education Partnership Initiative (MEPI) demonstrated that institutional support, faculty involvement, technical support, infrastructure, and user engagement are necessary for a sustainable e-learning system~\cite{vovides2014systems}. Liberia’s College of Health and Life Sciences (COHLS) project further emphasizes the need for local technical capacity, policy frameworks, and resilience to electricity and bandwidth disruption~\cite{walsh2018elearning}. Our architecture aligns with these principles by integrating robust backend services, facilitating institutional tagging, and offline-capable features.

\section{Methods}
Juvenotes, an AI-powered medical educational platform, was developed through a multi-faceted approach with emphasis on automation and user experience (UX) tailored for Kenyan medical schools. The platform’s design and implementation were informed by established challenges in medical education in LMIC and AI advancements.

\subsection{System Architecture}
Juvenotes employs a containerized microservices architecture to ensure portability, resilience, and independent scalability of its components. The system consists of two primary services:

\begin{itemize}[leftmargin=*, itemsep=0pt]
    \item \textbf{Web service:} This acts as the user interface, built with a Vue/TypeScript frontend and an AdonisJS backend. It manages the user interface, user sessions, institutional affiliations, and routes API calls. It also implements the database schema and business logic, such as mapping generated questions to specific courses and concepts.
    \item \textbf{AI engine:} A Python-based backend service that exposes a WebSocket API. Its primary function is to orchestrate calls to Azure cloud services (OCR and LLM) for document analysis and to format the results.
\end{itemize}

Both services are packaged as Docker containers and deployed on Ubuntu Virtual Machines (VMs) in Azure, with Coolify automating builds and orchestration. Services integrated into the architecture include MeiliSearch and Algolia for search indexing, Cloudinary for media hosting, PostHog for user analytics, Mailgun for email notifications, and M-Pesa for mobile payments. All inter-service communication is secured over internal networks within Azure.

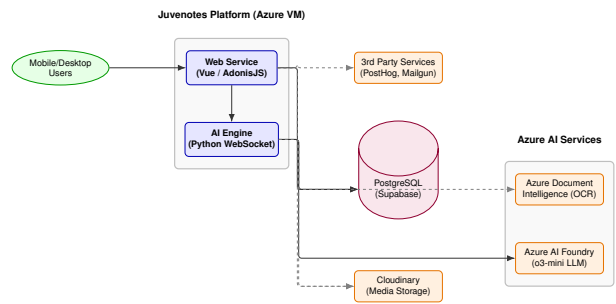
\begin{figure}[ht]
\centering
\resizebox{\columnwidth}{!}{
\begin{tikzpicture}
    \begin{scope}[node distance=2.1cm and 3.3cm]
        \node[user_node] (user) {Mobile/Desktop\\Users};

        \node[block, right=2.7cm of user] (web) {Web Service\\(Vue / AdonisJS)};
        \node[block, below=1.3cm of web] (ai) {AI Engine\\(Python WebSocket)};
        \begin{pgfonlayer}{background}
            \node[group_box, fit=(web) (ai), label={[yshift=0.5cm]above:\sffamily\bfseries Juvenotes Platform (Azure VM)}] {};
        \end{pgfonlayer}

        \node[cloud_service, right=2.7cm of web] (utils) {3rd Party Services\\(PostHog, Mailgun)};
        \node[data_store, below=1.3cm of utils] (db) {PostgreSQL\\(Supabase)};
        \node[cloud_service, below=1.8cm of db] (media) {Cloudinary\\(Media Storage)};

        \node[cloud_service, right=2.7cm of db] (ocr) {Azure Document\\Intelligence (OCR)};
        \node[cloud_service, below=1.3cm of ocr] (llm) {Azure AI Foundry\\(o3-mini LLM)};
        \begin{pgfonlayer}{background}
            \node[group_box, fit=(ocr) (llm), label={[yshift=0.5cm]above:\sffamily\bfseries Azure AI Services}] {};
        \end{pgfonlayer}
    \end{scope}

    \draw[arrow] (user) -- (web);
    \draw[arrow] (web) -- (ai);
    \draw[dashed_arrow] (web.east) -- ++(0.7,0) |- (utils.west);
    \draw[arrow] (web.east) -- ++(0.7,0) |- (db.west);
    \draw[dashed_arrow] (web.east) -- ++(0.7,0) |- (media.west);

    \draw[arrow] (ai.east) -- ++(0.7,0) |- (db.west);
    \draw[dashed_arrow] (ai.east) -- ++(0.7,0) |- (ocr.west);
    \draw[arrow] (ai.east) -- ++(0.7,0) |- (llm.west);

    \draw[dashed_arrow] (ai.east) -- ++(0.7,0) |- (media.west);
\end{tikzpicture}
}

\caption{Juvenotes infrastructure overview. The architecture separates user interface, processing, storage, third-party, and AI cloud services, with all connections routed for clarity.}
\end{figure}

\subsection{Data Flow \& Content Generation}
The core functionality of Juvenotes is the automated transformation of academic documents into structured question banks. The data flow is orchestrated through a sequence of steps:

\begin{enumerate}[leftmargin=*, itemsep=0pt]
    \item \textbf{Document Upload:} Users initiate the process by uploading educational documents, such as PDF files of past exam papers, via the Juvenotes web interface.
    \item \textbf{WebSocket Transfer:} The uploaded document is sent to the AI engine backend over a persistent WebSocket connection for real-time processing.
    \item \textbf{OCR Processing:} Upon receipt, the AI engine leverages Azure Document Intelligence to perform OCR, extracting structured text, including passages, tables, and layout information.
    \item \textbf{Question/Answer Generation:} The extracted text is subsequently fed into Azure AI Foundry (OpenAI o3-mini model, version 2025-01-31). This LLM identifies exam-style question-answer pairs (e.g., multiple-choice, short-answer) from the content and generates high-quality answers, formatting outputs for the platform's database schema.
    \item \textbf{Result Integration:} The generated Q\&A items are parsed and inserted into the PostgreSQL database, creating entries in relevant tables such as questions, mcq\_choices, saq\_parts, and past\_papers. 
    \item \textbf{User Interaction and Feedback:} Once in the database, the new questions are immediately accessible to students via the web UI. Student responses and feedback (e.g., correct/incorrect, ratings) are recorded in engagement tables for analytics and continuous improvement. 
\end{enumerate}

The Python backend orchestrates these steps, submitting API calls to OCR and LLM services, receiving responses, and transforming them into SQL insertions. Original documents and intermediate outputs (like OCR JSON) are also stored in Azure blob storage for persistence and auditing.

\subsection{UX Refinements}
Recognizing the unique infrastructural challenges in Africa, such as limited broadband and unreliable internet, Juvenotes was designed with adaptations:

\begin{itemize}[leftmargin=*, itemsep=0pt]
    \item Mobile-first responsive design for low-cost devices and varying screen sizes.
    \item Bandwidth optimization for fast, lightweight experience even with slow or unreliable internet.
    \item Seamless uploads with progress indicators, showing real-time OCR/AI processing logs.
    \item Localization and institutional tagging so questions can be filtered by school and course code; supports local context in AI prompts (e.g., Kenyan guidelines).
\end{itemize}

Pilot users reported that these mobile-friendly design and responsive feedback features significantly increased satisfaction.

\subsection{Deployment \& Operations}
\textbf{CI/CD:} Development occurs on a trunk branch, with merges to the web branch triggering Coolify to rebuild and redeploy Docker containers. This pipeline ensures consistent, versioned releases.

\textbf{Networking:} All services run on Azure VMs within a Coolify-managed private network. Internal DNS/service discovery allows secure container communication.

\textbf{Database:} PostgreSQL stores user data, content, and analytics. The schema supports a relational data model (see Appendix for details).

\textbf{Scalability:} The microservices architecture allows scaling of specific parts independently (e.g., OCR/LLM services). Coolify manages orchestration for resilience and elasticity.

\subsection{Database Schema Highlights}
The PostgreSQL schema organizes platform data into logical domains:

\begin{itemize}[leftmargin=*, itemsep=0pt]
    \item \textbf{User Management:} \texttt{users}, \texttt{roles}, and mappings for authentication and authorization.
    \item \textbf{Educational Content:} \texttt{mcq\_choices}, \texttt{concepts}, \texttt{courses}, \texttt{past\_papers}, \texttt{questions}, \texttt{saq\_parts}. Each question links to a past paper, course, and concept; MCQ tables store answer options.
    \item \textbf{Engagement Tracking:} \texttt{user\_mcq\_responses}, \texttt{user\_saq\_responses}, \texttt{user\_study\_sessions}, \texttt{analytics}, and \texttt{user\_study\_times}.
    \item \textbf{Feedback \& Progress:} \texttt{question\_feedbacks} (student ratings/comments), \texttt{user\_concept\_progress} (topic mastery).
    \item \textbf{Institution Integration:} \texttt{institutions}, \texttt{institution\_courses} mapping users/content to colleges/programs.
\end{itemize}

\begin{figure}[ht]
\centering
\resizebox{\columnwidth}{!}{
\begin{tikzpicture}
    \begin{scope}[node distance=1.3cm and 1.2cm]
        \node[table] (users) {\textbf{users}};
        \node[table, right=of users] (roles) {roles};
        \node[table, right=of roles] (institutions) {\textbf{institutions}};
        \node[table, right=of institutions] (inst_courses) {institution\_courses};
        \node[table, below=of inst_courses] (courses) {\textbf{courses}};
        \node[table, left=of courses] (past_papers) {\textbf{past\_papers}};
        \node[table, left=of past_papers] (concepts) {concepts};
        \node[table, below=of past_papers] (questions) {\textbf{questions}};
        \node[table, below left=of questions] (mcq_choices) {mcq\_choices};
        \node[table, below right=of questions] (saq_parts) {saq\_parts};
        \node[table, below=of mcq_choices] (user_mcq_resp) {user\_mcq\_responses};
        \node[table, below=of saq_parts] (user_saq_resp) {user\_saq\_responses};
        \begin{pgfonlayer}{background}
            \node[group_box, fit=(users) (roles) (institutions) (inst_courses), 
                  label=above:{\footnotesize \sffamily User \& Institution Management}] {};
            \node[group_box, fit=(past_papers) (concepts) (courses) (questions), 
                  label=above:{\footnotesize \sffamily Content Structure}] {};
            \node[group_box, fit=(mcq_choices) (saq_parts) (user_mcq_resp) (user_saq_resp), 
                  label=above:{\footnotesize \sffamily Engagement \& Responses}] {};
        \end{pgfonlayer}
    \end{scope}

    \draw[rel] (users.east) -- (roles.west);
    \draw[rel] (inst_courses.west) -- (institutions.east);
    \draw[rel] (inst_courses.south) -- (courses.north);
    \draw[rel] (past_papers.south) -- (questions.north);
    \draw[rel] (concepts.south) -- (questions.north);
    \draw[rel] (courses.south) -- (questions.north);
    \draw[rel] (questions.south) -- (mcq_choices.north);
    \draw[rel] (questions.south) -- (saq_parts.north);
    \draw[rel] (mcq_choices.south) -- (user_mcq_resp.north);
    \draw[rel] (saq_parts.south) -- (user_saq_resp.north);
\end{tikzpicture}
}
\caption{Simplified relational schema of the Juvenotes Supabase/PostgreSQL database, highlighting key user, content, and engagement tables.}
\end{figure}
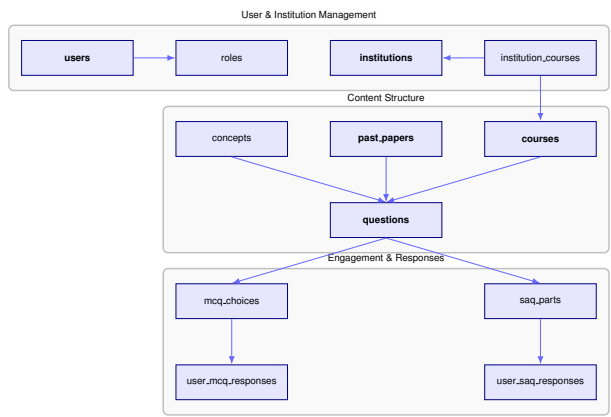

These normalized tables enable quick lookup of questions by topic or paper, and link user performance to learning objectives. The design emphasizes flexibility (e.g., multi-concept tagging) and performance (indexed search, stats caching).

\subsection{AI/ML Pipeline}
Azure Document Intelligence Studio performs OCR on uploaded documents, extracting text, key-value pairs, tables, and structures. The Read OCR model processes scanned exam papers, yielding JSON output of paragraphs, lines, and words, plus tables. For specialized formats, custom models can be trained to improve accuracy. The raw text/structure is then forwarded to the NLP stage.

For NLP and question generation, Azure AI Foundry (with OpenAI o3-mini, version 2025-01-31) is used. The OCR text is fed into this model with prompts to identify exam-style questions and generate answers. The model formats output (question, choices, correct answer, explanation) for the database schema.

The Python backend submits OCR/LLM API calls, receives responses, and transforms them into SQL insertions. Processed results and metadata (confidence, timestamps) are stored for auditing and analytics. Each generated question/source paper and model’s confidence/user feedback are recorded. Original documents and intermediate outputs (e.g., OCR JSON) are saved in Azure blob/object storage. Once in the DB, questions are immediately available to students.

\begin{figure}[ht]
\centering
\resizebox{\columnwidth}{!}{
\begin{tikzpicture}
    \begin{scope}[node distance=2.8cm]
        \node[block] (frontend) {Frontend\\(Vue/TypeScript)};
        \node[block, right=of frontend] (backend) {API Backend\\(AdonisJS)};
        \node[block, above right=1.2cm and 2.0cm of backend] (ai_engine) {AI Engine\\(Python)};
        \node[data_store, below right=1.2cm and 2.0cm of backend] (db) {Database\\(PostgreSQL)};
        \node[cloud_service, right=of ai_engine] (azure_ai) {Azure AI\\Services (OCR/LLM)};
    \end{scope}
    \draw[arrow] (frontend) -- (backend);
    \draw[arrow] (backend) -- (ai_engine);
    \draw[arrow] (backend) -- (db);
    \draw[arrow] (ai_engine) -- (db);
    \draw[arrow] (ai_engine) -- (azure_ai);
\end{tikzpicture}
}

\caption{Juvenotes internal microservices and data flow.}
\end{figure}
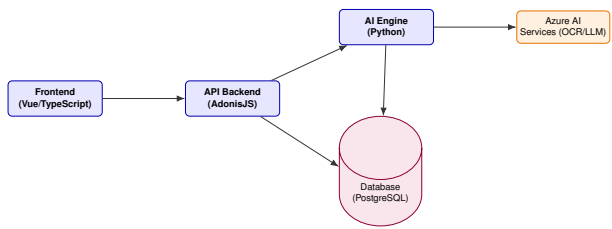

\subsection{Pilot Program and Evaluation}
Juvenotes was piloted over seven months at five Kenyan institutions: University of Nairobi, Moi University, Kenyatta University, Egerton University, and KMTC. Both medical students and faculty participated, uploading past papers and using the generated banks.

Evaluation included usage analytics and user surveys. E-learning interventions in LMICs have faced methodological challenges, often relying on subjective measures and custom frameworks, limiting comparability and validity~\cite{barteit2020evaluation}. Most prior studies were small-scale pilots and used summative assessments and questionnaires; knowledge testing was also frequent. Despite most reporting effectiveness, quality ratings showed low scientific rigor, especially for comparability/validity.

For Juvenotes, pilot data indicated:
\begin{itemize}[leftmargin=*, itemsep=0pt]
    \item \textbf{Increased Engagement:} 40\% rise in daily active users.
    \item \textbf{Faster Curation:} Content processing time dropped from days to minutes.
    \item \textbf{High Satisfaction:} 90\% of students reported higher satisfaction; instant access to questions and immediate feedback were valued.
\end{itemize}
Challenges included persistent connectivity, occasional AI question errors, and need for ongoing faculty review and correction. These underscore the importance of human oversight and robust offline design.

\subsection{Ethical, Social, Local Context}
Juvenotes was built with ethical and contextual awareness. Data privacy is prioritized; all data is encrypted in transit and at rest, complying with Kenyan and international standards. No personal data is used by AI services beyond academic necessity.

The platform is free for partner institutions, optimized for low-cost devices, and accommodates accessibility needs (large fonts, high contrast, language support). Student feedback is logged to fine-tune prompts and correct errors; faculty can flag inappropriate content before republishing (human-in-the-loop). This minimizes AI bias and misinformation and promotes equal access.

\section{Discussion}
Juvenotes highlights the transformative capacity of AI-driven platforms in addressing educational deficits in low-resource African settings. Its automated pipeline, converting academic documents to question banks, increased access and engagement. However, infrastructural and evaluative challenges remain, indicating a need for continued improvements and systemic approaches.

The platform was designed to tackle the WHO’s projected 6.1 million health-worker shortfall by automating content creation through Azure OCR and LLMs, addressing manual authoring bottlenecks. In the Kenyan pilot, daily active users increased by 40\%, content processing time dropped from days to minutes, and 90\% of students reported higher satisfaction. These results align with findings that improved user experience boosts satisfaction and engagement, even if measured knowledge gains are not yet statistically superior to traditional methods.

Design refinements addressed LMIC structural challenges. With broadband access at 36\%, mobile-responsive design, bandwidth optimization, and real-time upload feedback were essential~\cite{worldbank2023digital}. The containerized microservices architecture, deployed on Azure, provided scalability and resilience, addressing the “pilotitis” of under-scaled e-learning projects~\cite{olum2020medical}.

Despite successes, persistent challenges included rural connectivity and unreliable electricity. Occasional AI question errors highlight LLM “hallucination” issues, especially when not fine-tuned for local context~\cite{huang2023hallucination}. Human-in-the-loop review remains essential.

Sustainability requires long-term institutional investment, human resources, and formal curricular integration. The pilot was short-term, relying on subjective metrics and lacking rigorous, long-term evaluations. This matches reviews noting reliance on pilots, self-reported outcomes, and limited use of controlled, longitudinal studies and frameworks like Kirkpatrick’s model~\cite{barteit2020evaluation}.

To enhance Juvenotes and similar systems: fine-tuning models with local curricula and educator-led workflows will improve content quality. Empowering local content creators and enabling translations can increase relevance. Technological enhancements—offline access, intermittent sync, deeper institutional embedding—will help move beyond pilot status. Rigorous evaluation frameworks (pre/post-testing, quasi-experiments, Kirkpatrick Levels 3/4) will better measure impact.

\section{Conclusion}
Juvenotes demonstrates the potential of AI to democratize medical education in resource-limited environments by rapidly generating practice content and improving accessibility. To realize this potential, future efforts must combine technological innovation with human oversight, local contextualization, systemic integration, and methodological rigor in evaluation. Only through such a multi-faceted approach can AI-powered e-learning effectively strengthen healthcare systems and advance global health education goals.

\section*{Code Availability }
The source code for the entire Juvenotes web application (\texttt{juvenotes/web}) is available from the authors upon reasonable request.

\bibliographystyle{unsrt}

\end{document}